\documentclass[letter,seceq,twocolumn]{jpsj2}
\usepackage{txfonts}
\usepackage{ulem}
\newcommand{\kY}[3]{Y_{#1}^{#2}(\vtilde{#3})}

\newcommand{\ket}[1]{\left| #1  \right\rangle}
\newcommand{\ave}[1]{\left\langle  #1  \right\rangle}
\newcommand{\kvec}[1]{\boldsymbol{#1}}
\newcommand{\vhat}[1]{\kvec{\hat{ #1}}}
\newcommand{\vtilde}[1]{\kvec{\tilde{ #1}}}

\newcommand{\bracketi}[2]{\left\langle {#1}|{#2} \right\rangle}
\def\bracketii#1#2#3{\left\langle #1 \right|#2 \left|#3 \right\rangle}
\newcommand{\CG}[2]{\left( #1 | #2 \right)}
\newcommand{\threej}[6]{\left(
  \begin{array}{ccc}
    #1   & #2   & #3   \\
    #4   & #5   & #6   \\
  \end{array}
\right)}

\title{%
Theory of  Coupled Multipole Moments Probed by X-ray Scattering in CeB$_6$ 
}

\author{%
Hiroshi N. \textsc{Kono}\thanks{E-mail address: kono@cmpt.phys.tohoku.ac.jp}, Katsunori \textsc{Kubo}\thanks{Present address: Advanced Science Research Center, Japan Atomic Energy Research Institute, Tokai, Ibaraki 319-1195.
} and Yoshio \textsc{Kuramoto} 
}

\inst{%
Department of Physics, Tohoku University, Sendai 980-8578
}

\recdate{\today}

\abst{
A minimal model for multipole orders in CeB$_6$ shows that degeneracy of the quadrupole order parameters and strong spin-orbit coupling lead to peculiar temperature and magnetic-field dependences of the X-ray reflection intensity at superlattice Bragg points.  Furthermore, the intensity depends sensitively on the surface direction.  These theoretical results explain naturally recent X-ray experiments in phases II and III of CeB$_6$.  It is predicted that under weak magnetic field perpendicular to the (111) surface, the reflection intensity should change non-monotonically as a function of temperature.
}

\kword{%
X-ray scattering, $\text{CeB}_6$, quadrupole order, multipole order 
}

\begin{document}
\sloppy
\maketitle

In  $\mathrm{CeB_6}$ and related systems,
presence of both orbital and magnetic degrees of freedom brings about rich structures in the phase diagram.
In zero field, $\mathrm{CeB_6}$ turns into the antiferro quadrupole (AFQ) ordered  phase (phase II) at $T_\mathrm{Q} = 3.4$K
 and  turns into antiferromagnetic (AFM) ordered  phase (phase III) at $T_\mathrm{N}=2.3$K\cite{Effantin}.
The order parameter in phase II is the $\Gamma_{5g}$-type quadrupole moment.
In phase III, the non-collinear magnetic structure is described by four wave numbers:
$\kvec{k}_1=[1/4,1/4,1/2], \kvec{k}_1'=[1/4,-1/4,1/2]$, 
$\kvec{k}_2=[1/4,1/4,0], \kvec{k}_2'=[1/4,-1/4,0]$.

Recently, X-ray scattering has been utilized as a powerful probe
to detect orbital orderings\cite{murakami}.
For $\mathrm{CeB_6}$,
Yakhou \textit{et al}. have performed resonant and non-resonant scattering experiments and found superlattice reflections in phases II and III.\cite{Xray00}
Nakao \textit{et al}. have identified the boundary between phases I and II in a magnetic field by resonant scattering.\cite{Xray1}
More recently, Tanaka \textit{et al.} 
have reported unexpected temperature and magnetic field dependences using non-resonant X-ray scattering.\cite{Xray2,Xray3}
Although the superlattice spots emerge below $T_\mathrm{Q}$,
the intensity of $(n/2,n/2,n/2)$ reflections with $n$ odd integers remains small in $T_\mathrm{N}< T < T_\mathrm{Q}$. 
The intensity increases almost stepwise below $T_\mathrm{N}$.\cite{Xray2,Xray3}
Furthermore the small intensity in phase II is suppressed by application of a magnetic field of as small as 0.1 T, 
while the suppression in phase III requires an order of magnitude larger magnetic field\cite{Xray3}.
These features seem strange at first sight since the staggered quadrupole moments probed by X-rays should already be present in phase II, and should
not change significantly below the N\'{e}el temperature $T_\mathrm{N}$.
In this paper we demonstrate how X-rays probe the coupling 
among dipole, quadrupole and octupole moments, 
and show that consideration of a quasi-continuous symmetry of the quadrupole order parameters provides a  natural explanation of the experimental observations.

In $\mathrm{CeB_6}$, a localized 4f electron of a Ce$^{3+}$ ion has
the quartet crystalline electric field (CEF) ground state, which is called $\Gamma_8$ and is well below the excited CEF level $\Gamma_7$.\cite{CeB6}
An orbital pair in the $\Gamma_{8}$ level are given by
\begin{align}
\ket{ + \uparrow}=\sqrt{\frac{5}{6}}\ket{\frac{5}{2}}+\sqrt{\frac{1}{6}}\ket{-\frac{3}{2}}, \ 
\ket{- \uparrow }=\ket{ \frac{1}{2}},
\label{orbitals}
\end{align}
in terms of eigenstates of 
$J_z$.
The Kramers partners of $\ket{ \pm \uparrow}$ are written as
$\ket{ \pm \downarrow}$, and are obtained by reversing the sign of $J_z$ in eq. (\ref{orbitals}).
To describe multipole operators, we introduce two kinds of pseudo spin operators 
$\kvec{\sigma}$ and $\kvec{\tau}$ as
\begin{align}
\tau^z \ket{\pm \uparrow} =\pm \ket{\pm \uparrow}, \ \
\tau^z \ket{\pm \downarrow} =\pm \ket{\pm \downarrow},\nonumber \\
\sigma^z \ket{\pm \uparrow} = \ket{\pm \uparrow}, \ \
\sigma^z \ket{\pm \downarrow} =-\ket{\pm \downarrow}.
\end{align}
Then the magnetic moment $\kvec{M}$ is given by
\begin{align}
\kvec{M}=   \mu _\mathrm{B} \sum_i \left( 
\kvec{\sigma}_i +\frac 47 \kvec{\eta}_i \right), 
\label{moment}
\end{align}
where $\kvec{\eta}_i$ describes the orbital dependent part, and is given by
\begin{align}
\kvec{\eta} &=(\eta^+ \sigma^x, \eta^- \sigma^y, \tau^z \sigma^z), 
\end{align}
with
$\eta^{\pm} =(\pm\sqrt{3}\tau^x - \tau^z$)/2.
The $\Gamma_{5g}$-type quadrupole moment has components
\begin{align}
O_{yz} = \tau ^y \sigma^x, \ \ 
O_{zx} = \tau ^y \sigma^y, \ \ 
O_{xy} = \tau ^y \sigma^z.  
\end{align}
It is convenient to introduce a vector 
$\kvec{\mu} = (O_{yz},O_{zx},O_{xy})$.
The octupole moment with the $\Gamma_{5u}$ symmetry has also three components given by
\begin{align}
\kvec{\zeta} =(\zeta^+ \sigma^x, \zeta^- \sigma^y, \tau^x \sigma^z),
\end{align}
with
$\zeta^\pm =-(\tau^x\pm\sqrt{3}\tau^z)/2$.

We work with a RKKY type multipole Hamiltonian \cite{Ohkawa,phase2MF,KK} which reproduces phases II and III with minimum number of interactions.
The model under magnetic field $\kvec{H}$ is given by
\begin{align}
\mathcal{H} &=\sum_{\langle i j \rangle}\left( D_{5g}  \kvec{\mu}_i \cdot \kvec{\mu} _j +D_{4u2} \kvec{\eta}_i  \cdot \kvec{\eta}_j +D_{5u} \kvec{\zeta}_i \cdot \kvec{\zeta}_j \right)  \nonumber \\
&+ \sum _{\{i j\} } \sum_{ \gamma  \gamma ' }  K_{4u2}^{ \gamma  \gamma '} {\eta}_i^{\gamma}  {\eta}_j^{\gamma'} - \kvec{M}\cdot \kvec{H} +H_\mathrm{s},
\label{model}
\end{align}
where $\langle i j \rangle$ denotes a nearest neighbor pair, and $\{i  j\}$ denotes a next-nearest neighbor pair.
We have introduced
the  next-nearest interaction of the pseudo-dipole type: 
\begin{align}
K_{4u2}^{\gamma ' \gamma} = K_{4u2} (\delta^{\gamma, \gamma '} -3 n_{i j}^\gamma n_{i j }^ {\gamma'} )/12,
\end {align}
where $\kvec{n}_{i j }$ is the unit vector across the next-nearest neighboring sites  $i$ and $j$.\cite{KK}
This interaction stabilizes phase III with the 
peculiar pattern of dipole moments. 
For simplicity, we do not consider such part of dipole interactions that comes from $\kvec{\sigma}_i$ in eq. (\ref{moment}), nor the $\Gamma _{5u}$ type pseudo-dipole interaction, which is known to be important for realizing phase III' in magnetic field larger than 1 T\cite{KK}.  Hence discussion of phase III' is out of the scope of this paper.
We assume that $D_{5g}$ is positive and is the largest among all interactions.
Then the $\Gamma_{5g}$-type AFQ order first sets in from the paramagnetic phase. 
The term $H_\mathrm{s}$ is introduced to simulate the symmetry breaking due to the surface, as explained later. 

We study this Hamiltonian by the mean field theory
with superlattice structures up to $\sqrt{8}\times\sqrt{8}\times 2$ supercell, 
which can describe the magnetic structure in phase III.
We take the energy unit as the quadrupole ordering temperature: $T_\mathrm{Q} = 1$ or, equivalently, $D_{5g}=1/6$.
With the choice $D_{4u2}=D_{5u}=0.9 D_{5g}$ and
$K_{4u2}=0.5$, 
we obtain the magnetic ordering  temperature $T_\mathrm{N} =0.47$ with zero field.
Our mean field theory indeed stabilizes the 
pattern of the dipole moments proposed in ref.\citen{Effantin}.

In phase II, this model realizes the $\Gamma_{5g}$-type order parameter with three components $(O_{yz}, O_{zx}, O_{xy})$ at each site. 
In the mean field theory for eq. (\ref{model}),  there is no preferred direction for
$\kvec{\mu}$
without magnetic field.
We call this situation a quasi-continuous symmetry.
Because of the spin-orbit coupling, an external magnetic field tends to align $\kvec{\mu}\parallel \kvec{H}$.  This coupling is apparent in $\kvec{\eta}$ in eq. (\ref{moment}).
In other words, dipole and octupole moments combine to give $\kvec{\eta}$ under the cubic symmetry.\cite{phase2MF,KK}
In phase III, on the other hand, simultaneous presence of dipole and octupole orders restrict the direction of $\kvec{\mu}$ even without magnetic field.
In real CeB$_6$, there should be various symmetry breaking perturbations to fix the direction of $\kvec{\mu}$ even in phase II.   
We consider in particular such effect of surface that is simulated by the following term:
\begin{align}
H_\mathrm{s} = -E\sum_{i} \kvec{\mu }_i\cdot\vhat{\epsilon},
\label{surface}
\end{align} 
where $\vhat{\epsilon}$ is the unit vector normal to the surface, and 
 $E(>0)$ assumes that the surface prefers the wave function extended parallel to  it.
   For example, the (0,0,1) surface prefers $O_{xy}$ to other components $O_{yz}, O_{zx}$.
In phase II where AFQ is present, the (0,0,1) surface disfavors the 
AFQ component $\langle O_{xy}\rangle_{\rm AFQ}$ of the order parameter.
This situation is analogous to the N\'{e}el state in the 
antiferromagnetic Heisenberg model, where a magnetic field disfavors 
the magnetic moment parallel to the field.
Since we do not go into the details of the surface region, 
we take the summation over $i$ in eq. (\ref{surface}) for the whole system.

We now consider the consequences of symmetry breaking by the surface and by magnetic field on X-ray reflection.
We take two typical conditions as shown in Table \ref{table1}, which correspond  to recent experimental configurations \cite{Xray2,Xray3}.
\begin{table}[td] 
\caption{Typical directions of surface and magnetic field.}
\begin{center} 
\begin{tabular}{ccc}
\hline configuration 
& S(111) & S(110) \\
\hline
surface direction &  $(1,1,1)$& $(1,1,0)$\\
magnetic field direction &  $(\bar{2},1,1)$&$(\bar{1},1,0)$\\
$(hkl)$& $(n/2, n/2, n/2)$ & $(h/2, h/2, 1/2)$\\
& $n=1,3,5,7,9,11$&$h=5,7,9,11$\\
\hline
\end{tabular} \end{center} \label{table1} \end{table}
To calculate the Thomson scattering intensity, 
we need the structure factor $F(\kvec{\kappa})$ defined by
\begin{equation}
F(\kvec{\kappa})= 
\sum _{n,a} {\rm e}^{\mathrm{i} \kvec{\kappa }\cdot \kvec{R}_n}
p_a \bracketii{a} {\exp (\mathrm{i} \kvec{\kappa }\cdot \kvec{r}) }{a},
\label{Fe}
\end{equation}
where $\kvec{R}_n$ specifies a Ce site, and the coordinate $\kvec{r}$ 
of a 4f electron is measured from the Ce site.
The momentum transfer is given by $\kvec{\kappa}$.
The state labeled $a$ has the statistical weight $p_a$.
Then the scattering cross section is given by
\begin{align}
\left(  
{\frac{\mathrm{d} \sigma }{\mathrm{d} \Omega }}\right)_{\kvec{\epsilon}\rightarrow \kvec{\epsilon '}}=
\bigl|
\frac{e^2}{mc^2} F(\kvec{\kappa})
\kvec{\epsilon} \cdot \kvec{\epsilon'} \bigr|^2,
\end{align}
where $\kvec{\epsilon}$ 
denotes polarization of the incident beam 
and 
$\kvec{\epsilon '}$ that of the scattered beam.\cite{Xray0}

We expand $\ket{a}$ in terms of the total angular momentum basis $\ket{J M}$ with $J=5/2$ as
$\ket{a}=\sum _M c_M \ket{J M}$.
Then we obtain in eq. (\ref{Fe}) 
\begin{align}
&\bracketii{a}
{\exp (\mathrm{i} \kvec{\kappa}\cdot \kvec{r})}
{a}\nonumber \\
=&\sum _{M M'}c_M^\ast  c_{M'} \sum _{m,m'}\sum _{m_s}
\bracketii{lm}{\exp (\mathrm{i} \kvec{\kappa }\cdot \kvec{r})}{lm'} \nonumber \\
&\times \CG{JM}{lmsm_s}\CG{lm'sm_s}{JM'},
\end{align}
where $\CG{JM}{lmsm_s}$ is a Clebsch-Gordan coefficient with $l=3$ and $s=1/2$.
The wave function $\bracketi{\kvec{r}}{lm}$ is factorized into the radial and angular parts as
$\bracketi{\kvec{r}}{lm}=f(r)\kY{m}{l}{r }$, where $r=|\kvec{r}|$ and 
$\tilde {\kvec{r}}$ is the solid angle. 
From this we obtain
\begin{align}
&\bracketii{lm}{\exp (\mathrm{i} \kvec{\kappa }\cdot \kvec{r})}{lm'}  \nonumber \\
=&\sqrt{4\pi }\sum _{K=0}^{6}\ave{j_{K}(\kappa )}\mathrm{i}^{K}\sqrt{2K+1}(2l+1) \nonumber \\
&\times \sum _{Q=-K}^{K}(-1)^{Q}\kY{-Q}{K}{\kappa }\nonumber \\
&\times (-1)^m\threej{l}{K}{l}{0}{0}{0}\threej{l}{K}{l}{-m}{Q}{m'}.
\end{align}
Here we have introduced the notation $\ave{j_{K}(\kappa )}$ by
\begin{align}
\ave{j_{K}(\kappa )}=\int \mathrm{d} r j_{K}(\kappa r) f(r)r^2,
\end{align}
where $j_K(\kappa r)$ is a spherical Bessel function of order $K$.
We use the data of $\ave{j_K(\kappa)}$ calculated by Freeman and Desclaux with Dirac-Fock method\cite{Freeman}.

\begin{figure}
 \begin{center}
\includegraphics[width=8cm,keepaspectratio]{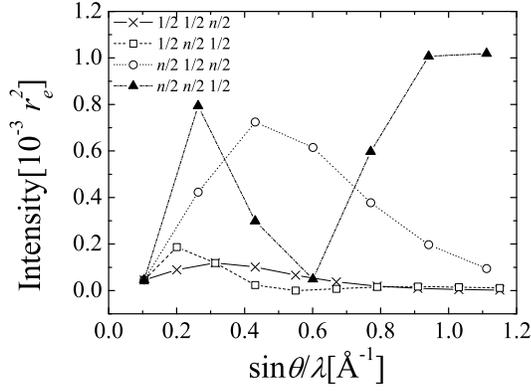}
 \caption{
The superlattice reflection intensities in the $O_{xy}$ AFQ state with $n$ odd integer against 
$\sin\theta / \lambda $.
 }
 \label{Oxy}
 \end{center}
\end{figure}

Before considering the effects of magnetic field and surface, 
we impose by hand an AFQ order with only
$O_{xy}$ and derive the X-ray scattering intensity.
This artifice helps us to identify the relationship between the AFQ order and $\kvec{\kappa}$.
The wave functions diagonalizing $O_{xy}$ are written as
\begin{align}
\ket{A \uparrow} &=\frac{1}{\sqrt{2}}
(\ket{+\uparrow }+\mathrm{i}\ket{-\uparrow }),\nonumber \\
\ket{B \uparrow} &=\frac{1}{\sqrt{2}}
(\ket{+\uparrow }-\mathrm{i}\ket{-\uparrow }),
\end{align}
and their Kramers partners $\ket{A \downarrow}$ and $\ket{B \downarrow}$.
The eigenvalue of $O_{xy}$ is $\pm 1$
for $\ket{A \uparrow\downarrow}$,
and $\mp 1$ for $\ket{B \uparrow\downarrow}$.
We take the staggered (G-type) AFQ order with complete occupation of
$\ket{A \uparrow}$ for the sublattice A, 
and that of $\ket{B \uparrow}$ for the sublattice B.
Since the X-ray intensity does not depend on the spin direction, this ferromagnetic configuration serves to analyze basic features of the intensity.
Figure \ref{Oxy} shows the intensities of superlattice reflections 
against $\sin \theta / \lambda = \kappa/(4\pi)$, where 
$\lambda$ is the wavelength of X-ray,
and $2\theta$ is the scattering angle.
The intensity is defined as the scattering cross section with $|\kvec{\epsilon}' \cdot \kvec{\epsilon}|=1$, and 
$r_e$ is the classical electron radius $r_e = e^2/mc^2$.
From Fig. \ref{Oxy}, it can be seen that the intensities have strong dependence on the direction of $\kvec{\kappa}$.  
The reason for this behavior is explained qualitatively as follows:
In the AFQ state, the structure factor with $\kvec{\kappa}$ being a superlattice vector
is written as 
$F(\kvec{\kappa}) \propto f_A(\kvec{\kappa})-f_B(\kvec{\kappa})$, 
where $f_{A(B)}(\kvec{\kappa})$ denotes the form factor of
the $A(B)$ sites.
There are high-symmetry directions from which 
A and B sublattices are seen as having the same projected charge density.
If a scattering vector is along such direction,
we obtain $F(\kvec{\kappa}) =0$ since $f_A(\kvec{\kappa})=f_B(\kvec{\kappa})$.
In the case of $O_{xy}$ AFQ state, such directions are $x$ and $y$ axes.
Therefore the $(1/2,n/2,1/2)$ intensity with large $n$ is small in Fig. \ref{Oxy}.
On the other hand,  we also have $f_A(\kvec{\kappa})=f_B(\kvec{\kappa})$ for $\kvec{\kappa}\parallel (0,0,1)$ since the different distribution of charge density in the $xy$ plane does not survive integration over this plane.  Thus the $(1/2,1/2,n/2)$ intensity is also small with large $n$.

To the contrary, the difference of the charge density should be seen most effectively from the direction parallel to $(1, 1, 0)$.  
In our calculation shown in Fig. \ref{Oxy}, 
$(n/2, n/2, 1/2)$ reflection has a deep minimum around $\kappa/{4\pi}\sim 0.6 {\rm \AA}^{-1}$.
This comes from interference of $K=2$ and $K=4$ contributions as discussed in ref.\citen{XrayT}.
Namely, we obtain
 $$
 F(\kappa) \propto \hat{\kappa}_x \hat{\kappa}_y \{ j_2(\kappa) +5/2(7 \hat{\kappa}_z^2 -1)  j_4(\kappa)\},
 $$
 where $\hat{\kappa}_\alpha$  is directional cosine along the $\alpha$ axis.\cite{XrayT,Xray4} 
Because $\hat{\kappa}_z$ becomes small as $\kappa$ increases in the $(n/2, n/2, 1/2)$ direction, the coefficient of $j_4$ becomes negative and suppresses the $K=2$ contribution.
Hence the scattering intensity becomes minimum when these two contributions nearly cancel each other.

We now proceed to mean-field treatment of phases II and III, and discuss
X-ray intensities without assuming the AFQ pattern a priori.
We tentatively take $E = 5 \times 10^{-3}$ in eq. (\ref{surface}).
\begin{figure}
\begin{center}
\includegraphics[width=8cm,keepaspectratio]{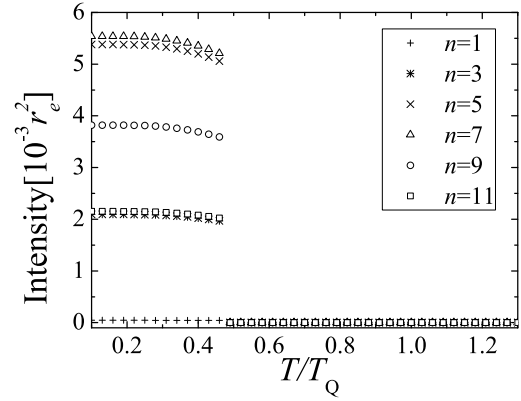}
\caption{The temperature dependence of X-ray intensity for the configuration S(111). The scattering vector is $(n/2, n/2, n/2)$.  The Intensity remains zero in phase II.}
 \label{XRAYEXPI}
 \end{center}
\end{figure}
Figure \ref{XRAYEXPI}  shows the temperature dependence of $(n/2, n/2, n/2)$ reflections in the configuration S(111).
It can be seen that $(n/2, n/2, n/2)$ reflections have no intensity in phase II. 
By the term simulating the surface effect, 
$\kvec{\mu}$ avoids the direction $(1, 1, 1)$ in phase II.
This means that the order parameter does not have a component 
$O_{yz}+O_{zx}+O_{xy}$, which alone contributes to
a finite scattering intensity for $(n/2, n/2, n/2)$ reflections. 
Other components lead to 
$f_A(\kvec{\kappa})=f_B(\kvec{\kappa})$, and
the intensity is zero even with a quadrupole order.

On entering phase III,  $\kvec{\mu}$ gains a component in the direction of $(1,1,1)$, 
since the surface term is no longer the only source to fix the direction of $\kvec{\mu}$. 
In the domain where we have the dipole order of $(1/4,\pm 1/4, 1/2)$,
the $O_{xy}$ component of 
quadrupole moment is stabilized by the spin-orbit coupling. 
Another component $O_{yz}$ is stabilized in the $(1/2,\pm 1/4, 1/4)$ domain, and
$O_{zx}$ in the $(1/4, 1/2, \pm 1/4)$ domain.
These domains give same superlattice intensity in the configuration S(111). 
In our mean-field theory,  dipole and octupole order parameters develop continuously from zero below $T_{\mathrm N}$.
However, the direction of $\kvec{\mu}$ changes discontinuously at $T_{\mathrm N}$,
and the intensity changes discontinuously.
Actual experimental results show small but finite intensities of $(n/2, n/2, n/2)$ in phase II with the configuration S${(111)}$. \cite{Xray2}
We interpret this feature in terms of 
 slight mixture of unfavorable component $O_{yz}+O_{zx}+O_{xy}$ in 
the sample by imperfections other than surface
and by multidomain effects.

\begin{figure}
 \begin{center}
\includegraphics[width=8cm,keepaspectratio]{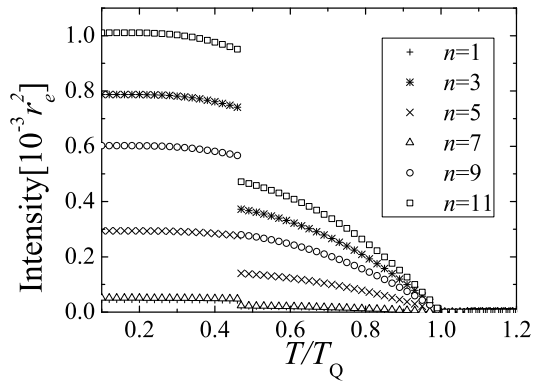}
 \caption{The temperature dependence of X-ray intensity in the  configuration S(110).  The scattering vector is $(n/2, n/2, 1/2)$. }
 \label{XRAYEXPII}
 \end{center}
\end{figure}

In the configuration S(110), the surface favors $\kvec{\mu}\perp (1,1,0)$.  
The X-ray scattering with 
$\kvec{\kappa}  \parallel (1,1,0)$
 probes the quadrupole component $O_{xy} =\mu_z$. 
Hence $\kvec{\mu}\parallel (0,0,1)$, which is favored by eq. (\ref{surface}), contributes to the scattering.
On the other hand, another favored component $\kvec{\mu}\parallel (\bar{1},1,0)$ is not probed by X-rays with $\kvec{\kappa}\parallel (1,1,0)$.  
In this paper, we take the simplest approach to take both quadrupole configurations into account. 
Namely, we choose $\kvec{\mu} \parallel (\bar{1}, 1, \sqrt{2})$ which has equal weights of  $(\bar{1},1,0)$ and $(0,0,1)$ configurations.
Figure \ref{XRAYEXPII} shows the temperature dependence of $(n/2, n/2, 1/2)$ reflections for the configuration
S(110).
The intensities of $(n/2, n/2, 1/2)$ reflections become finite below $T_\mathrm{Q}$ and increase as the temperature is decreased.
On entering phase III, the intensity increases discontinuously by the same reason in the case of S(111).

\begin{figure}
 \begin{center}
\includegraphics[width=8cm,keepaspectratio]{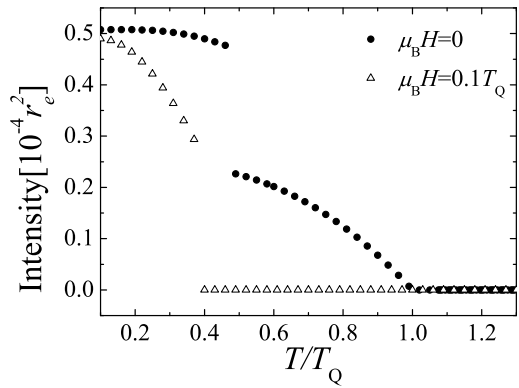}
 \caption{Temperature and magnetic field dependence of  the X-ray intensity in the configuration S(110). The scattering vector is $(7/2, 7/2, 1/2)$ and magnetic field is along $(\bar{1},1,0)$.}
  \label{g5m}
 \end{center}
\end{figure}

A magnetic field can also influence the direction of   $\kvec{\mu}$ by the spin-orbit interaction.
Figure \ref{g5m} shows the dependence of X-ray scattering intensities on magnetic field in the configuration S(110).
The intensity of $({7}/{2}, {7}/{2}, {1}/{2})$ reflection has a finite value without field, but is suppressed by a magnetic field.
This is because the magnetic field along $(\bar{1},1,0)$ 
rotates $\kvec{\mu}$ so as to be parallel to $\kvec{H}$.
Since $\kvec{\mu}$ in this direction is perpendicular to the scattering direction 
$({7}/{2}, {7}/{2}, {1}/{2})$, the intensity vanishes although the quadrupole order is present.

\begin{figure}
 \begin{center}
\includegraphics[width=9cm]{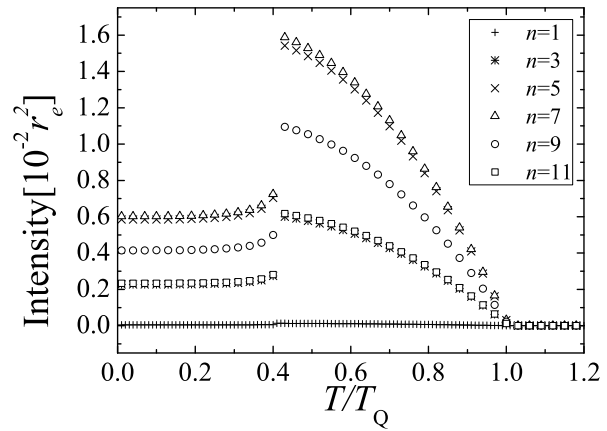}
 \caption{The temperature dependence of X-ray intensity in the configuration S(111) with $\kvec{H}\parallel (1,1,1)$ and $\mu_\mathrm{B}H = 0.1 T_\mathrm{Q}$.   
 The scattering vector is $(n/2, n/2, n/2)$. }
 \label{H//(111)}
 \end{center}
\end{figure}

In the actual result for the configuration S(111), 
a small intensity which is already present in phase II
 is suppressed by a small magnetic field \cite{Xray3}.
This is explained in terms of 
rotation of $\kvec{\mu}$ toward $\kvec{H}$ parallel to the surface.  
If the present mechanism is realistic,  magnetic field along $(1,1,1)$ should lead to {\it reduction} of superlattice intensity on entering phase III.    The result of model calculation is shown in Fig.  \ref{H//(111)}. 

To summarize, we have clarified the effects of quasi-continuous symmetry of the quadrupole moments on X-ray scattering from CeB$_6$.
The couplings with dipole and octupole moments lead to unexpected features; 
large increase of the superlattice intensity on entering phase III, and strong suppression of intensity by magnetic field.
Our theory provides an interpretation of recent experimental results.\cite{Xray2,Xray3}
If magnetic field is applied perpendicular to the $(1,1,1)$ surface, our theory predicts {\it increase} of the scattering intensity in phase II.  This is because the magnetic field induces the order parameter $O_{yz}+O_{zx}+O_{xy}$.
Since the dipole order in phase III will rotate $\kvec{\mu}$ from this preferred direction, the intensity should {\it decrease} on entering phase III.
We further predict that an intensity minimum of the superlattice reflection as a function of $\kappa$ should be observed if $(n/2, n/2, 1/2)$ reflections are measured in phase III with the $(1,1,0)$ surface.

We would like to thank Y. Tanaka and K. Katsumata for showing their experimental results prior to publication, 
and S.W. Lovesey, H. Kusunose, S. Ishihara and G. Sakurai for valuable discussion.
One of the authors (H. N. K. ) gratefully acknowledges useful comments by S. Suzuki.

\end{document}